\documentclass[onecolumn,english,prb,showpacs]{revtex4}
\usepackage[T1]{fontenc}
\usepackage[latin9]{inputenc}
\usepackage{amssymb}
\usepackage{graphicx}
\usepackage{amsmath,color}
\usepackage{mathrsfs}
\usepackage{float}
\usepackage{indentfirst}

\makeatletter

\def\journal #1, #2, #3, 1#4#5#6{{\sl #1~}{\bf #2}, #3 (1#4#5#6) }

\newcommand{\nn}{\nonumber\\}

\makeatother

\begin{document}

\title{Frequency domain winding number and interaction effect on topological insulators}

\author{Lei Wang$^{1,2}$, Xi Dai$^1$ and X. C. Xie$^{2,3}$}

\affiliation{$^{1}$Beijing National Lab for Condensed Matter Physics and Institute of Physics, Chinese Academy of Sciences, Beijing 100190, China }

\affiliation{$^{2}$International Center for Quantum Materials, Peking University, Beijing 100871, China}

\affiliation{$^{3}$Department of Physics, Oklahoma State University, Stillwater, Oklahoma 74078, USA}

\begin{abstract}
We study the effect of interactions on the time reversal invariant topological insulators in four and three spatial dimensions. Their topological indices are expressed by the interacting Green's functions. Under the local self-energy approximation, we find that interaction could induce nontrivial frequency-domain winding numbers and change the topological classes of the system. Our results suggest that the topological phases could be destroyed without developing long range orders. Practical issues on the accurate frequency-momentum integration combined with DMFT and diagrammatic calculations of the interacting Green's functions are also addressed.
\end{abstract}

\pacs{73.43.-f, 71.70.Ej, 03.65.Vf, 71.27.a}





\maketitle

\section{Introduction}
Recently, there has been a raising interest on a class of topological phase of matter: the topological insulators, see \cite{Hasan:2010p23520, Moore:2010p15238} for reviews. Questions like what is the interaction effect on the topological insulators and how to characterize the topological phase beyond single particle basis pose intriguing challenges. There are some previous studies on the interacting effect on topological insulators. \citet{Wang:2010p25724} performed exact diagonalization and quantum Monte Carlo studies of the interacting Haldane model. \citet{Hohenadler:2011p28789, Zheng:2010p24293,Yu:2011p35101} performed quantum Monte Carlo and variational cluster approach studies on the Kane-Mele-Hubbard (KMH) model. One interesting finding in these studies is that the topological phases give their way to a gapped featureless state (spin liquid phase for KMH model) at small value of the spin orbital coupling. The possibility of fluctuation effect driven by interaction breaks the topological phase significantly go beyond the mean-field treatments \cite{Rachel:2010p20458, Wang:2010p25724}. Since in the latter case, long range orders (LRO) are needed to compete with the topological phases.

However, one unsatisfactory feature in these studies is that there lacks a direct characterization of the topological order: the phase boundaries are mostly determined by indirect signatures such as the correlation functions for LRO or the excitation gaps, rather than a sharp change of a topological indices. This leaves the mechanism and the nature of the fluctuation induced topological transitions unclear.

In this paper, we study the interaction effects on the topological insulators based on their single-particle Green's functions\cite{Wang:2010p25171}. We start from the four dimensional Chern insulator \cite{Zhang:2001p35316} since it is considered as the mother state of large class of topological insulators \cite{Qi:2008p12545,Wang:2010p25171}. The system's topological index-- Chern number-- is expressed by the following formula:



\begin{eqnarray}
n  =  \frac{\varepsilon_{\mu \nu \rho\sigma\tau }}{15 \pi^2} \mathrm{Tr} \int_{\mathrm{BZ}}  \frac{\mathrm{d}^{4}k}{(2\pi)^{4}}  \int_{-\infty}^{\infty} \frac{\mathrm{d}\omega}{2\pi}\,
 G \partial_{\mu} G^{- 1} G \partial_{\nu} G^{- 1} G \partial_{\rho} G^{- 1}G \partial_{\sigma} G^{- 1}G \partial_{\tau} G^{- 1}
\label{eqn:ishikawa}
\end{eqnarray}


where $\varepsilon_{\mu\nu\rho\sigma\tau}$ is the five-order anti-symmetric tensor, $\mu, \nu...\tau$ indices denote the frequency-momentum $(\omega, k_x, k_y, k_{z}, k_{\lambda})$  indices. The frequency integration ranges from $-\infty$ to $\infty$ and the momentum integration is over the first Brillouin zone.

To the authors knowledge, in the quantum Hall effect context the formula was first appeared  in \cite{So:1985p29974} and \cite{ISHIKAWA:1986p24092}. It was latter used to study the finite temperature and disorder effect on integer quantum hall effect\cite{ISHIKAWA:1987p26035,IMAI:1990p26297}. It was also used extensively by Volovik in his book\cite{Volovik:2003p1862}. Recently, Ref.\cite{Qi:2008p12545,Wang:2010p25171} proposed to use it and its variants as the topological order parameters of the interacting topological insulators. It is also used to study the boundary effect of topological insulators by \cite{Gurarie:2011p28415, Essin:2011p30698}. Physically, the formula descries Hall conductance of the material since it appears as the coefficient of the Chern-Simons term. On the mathematical side, the outcome of the integration Eq.\ref{eqn:ishikawa} is guaranteed to be an integer with physical Green's functions for a gapped system since it represents the homotopy index $\pi_5(GL(M,\mathbb{C}))=\mathbb{Z}$.

For noninteracting models $G^{- 1} (\omega,\bold{k}) = i \omega - H_\bold{k}$. The frequency integration could be done analytically, results into the Chern number \cite{Qi:2008p12545} expressed by momentum space integrations, \textit{i.e.} the TKNN numbers\cite{Thouless:1982p24208}. For interacting case, the Green's function becomes $G^{-1}(\omega,\bold{k})=i\omega-H_{\bold{k}}-\Sigma(\omega, \bold{k})$. The self-energy $\Sigma(\omega, \bold{k})$ contains the information of many-body physics. The fluctuation effect of many body physics are encoded in the frequency dependence of the self-energy. In general, the five-fold frequency-momentum integration in Eqn.\ref{eqn:ishikawa} becomes involved and one can not complete the frequency integration analytically. In mean-field studies of the interaction models \cite{Raghu:2008p3865, Rachel:2010p20458} and Born-approximation study of the disorder effects on topological insulators \cite{Groth:2009p11689}, the self-energy is approximately frequency independent ($\Sigma(\omega,\bold{k})\approx \Sigma_{\bold k}$) and it is effect is effectively absorbed into $H_{\bold{k}}$ and modified the topological classes thereof. However, more ubiquitous and intriguing cases are the self-energy has strong frequency dependences where the many body effect does not simply renormalize the single particle Hamiltonians. The main goal of the present paper is to study how nontrivial frequency dependency of the self-energy affects the topological classes. In section \ref{sec:local} we introduce the local self-energy approximation and study its consequences on the interacting Chern number of the four-dimensional quantum Hall effect. In section \ref{sec:Z2} we study the effect of frequency domain winding number on the $Z_{2}$ index of the three-dimensional topological insulators. We discuss issues on the practical calculations of the interacting Green's function in section \ref{sec:practical}.

\section{Local Self-energy approximation and frequency domain winding numbers \label{sec:local}}

We assume the self-energy is local and show how the local fluctuation affects the topological properties. Under this assumption the self-energy is independent of momentum $\bold{k}$, \textit{i.e.} $\Sigma(\omega, \bold{k})\approx\Sigma(\omega)$ but preserves full frequency dependence \footnote[1]{Notice that in general the self-energy is a matrix on the local orbital basis. Here we assume it is diagonal and has no orbital dependence, this could be thought of as a generalized \emph{local} assumption.}. The local self-energy approximation has been widely adopted in the investigations of strongly-correlated systems under the name dynamical-mean-field theory (DMFT) \cite{Georges:1996p5571} and it was proofed to be an accurate approximation for high dimensions\cite{Metzner:1989p6797,Rozenberg:1992p1863}.

In the following we show that with nontrivial frequency dependent self-energy, the frequency-domain winding number (FDWN) could affect the topological class of the system. Specifically, the interaction induced fluctuation effect could change the topological classes without developing LRO.

Without loss of generality, we assume the Hamiltonian could be expanded on the Dirac gamma matrices $H_{\bold{k}} = \sum_{a=1}^{5} h^{a}_{\bold{k}} \Gamma^a$, $h_{\bold{k}}$ is a five component vector whose module is denoted as $|h_{\bold{k}}|$. We also introduce the normalized vector $\hat{h}^{a}_{\bold{k}}\equiv h^{a}_{\bold{k}}/|h_{\bold{k}}|$. Under the local self-energy assumption the Green's function could be written as $G^{- 1} =i\omega- h^{a}_{\bold{k}}\Gamma^{a}-\Sigma(\omega) \equiv G^{- 1}_{\mathrm{atom}} - h^{a}_{\bold{k}}\Gamma^{a}$, where we introduced the atomic Green's function $G^{-1}_\mathrm{atom} \equiv i\omega-\Sigma(\omega)$ which contains all of the frequency dependence. Several lines of algebra \cite{Qi:2008p12545} shows that the interacting Chern number is

\begin{widetext}

\begin{eqnarray}
  n = \frac{2}{\pi^{2}} \int_{\mathrm{BZ}} \mathrm{d}^4 k  \int_{-\infty}^{\infty} \frac{ \mathrm{d} \omega}{2
  \pi i} \frac{\partial_{\omega} G^{- 1}_\mathrm{atom}}{(G^{- 2}_\mathrm{atom} - |h_{\bold{k}}|^2)^3} \varepsilon_{abcde} h_{\bold{k}}^{a} \partial_{k_{x}}
  h_{\bold{k}}^{b} \partial_{k_{y}} h_{\bold{k}}^{c} \partial_{k_{z}} h_{\bold{k}}^{d} \partial_{k_{\lambda}} h_{\bold{k}}^{e}
\label{eqn:separate}
\end{eqnarray}

Denote $z=G_{\mathrm{atom}}^{- 1}(\omega)$ and notice that it defines a mapping from real axis (frequency $\omega$) to the complex plane, we could finish the frequency integration

 \begin{eqnarray}
  \int_{- \infty}^{\infty}  \frac{\mathrm{d}\omega}{2 \pi i}  {\frac{\partial_{\omega}
  G_{\mathrm{atom}}^{- 1}}{(G_{\mathrm{atom}}^{- 2} - |h_{\bold{k}}|^2)^3} } 
  & = &
  \int_{ z (-\infty)}^{z (\infty)} \frac{\mathrm{d}z}{2 \pi i}
  {\frac{1}{(z^2 - |h_{\bold{k}}|^2)^3}} \nn
    & = &
  \oint_{\mathcal{C}} \frac{\mathrm{d}z}{2 \pi i}
  {\frac{1}{(z^2 - |h_{\bold{k}}|^2)^3}} \nn
   & = & \frac{3}{16|h_{\bold{k}}|^{5}} \gamma
   \label{eqn:freq}
\end{eqnarray}

\end{widetext}

\begin{figure}[!htbp] 
   \centering
     \includegraphics[width=9cm]{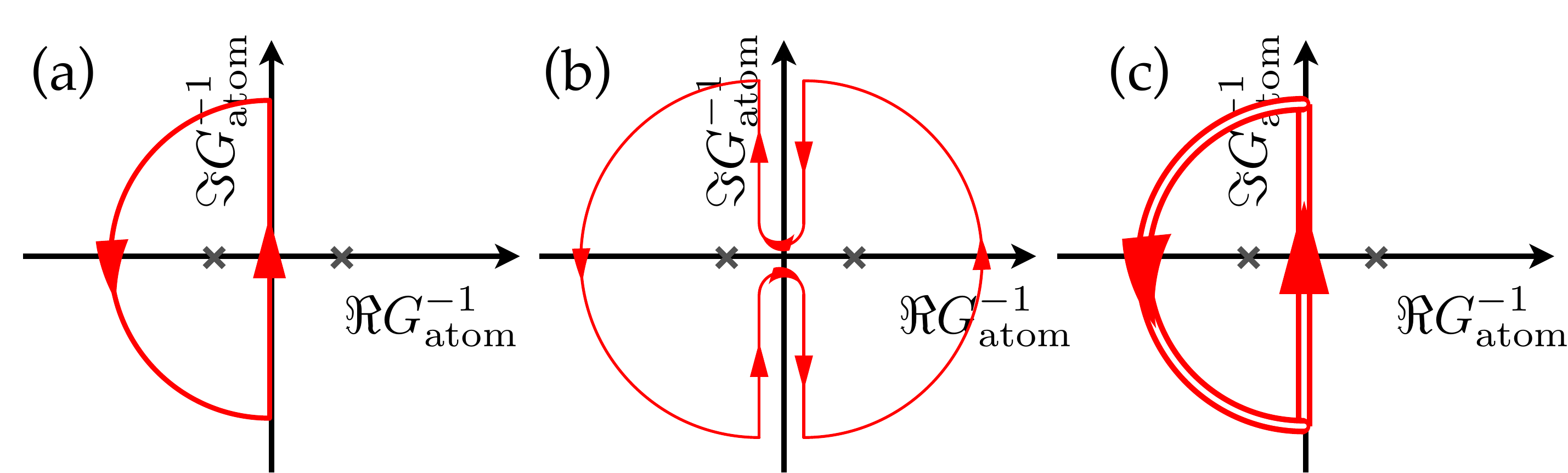}
     \includegraphics[width=9cm]{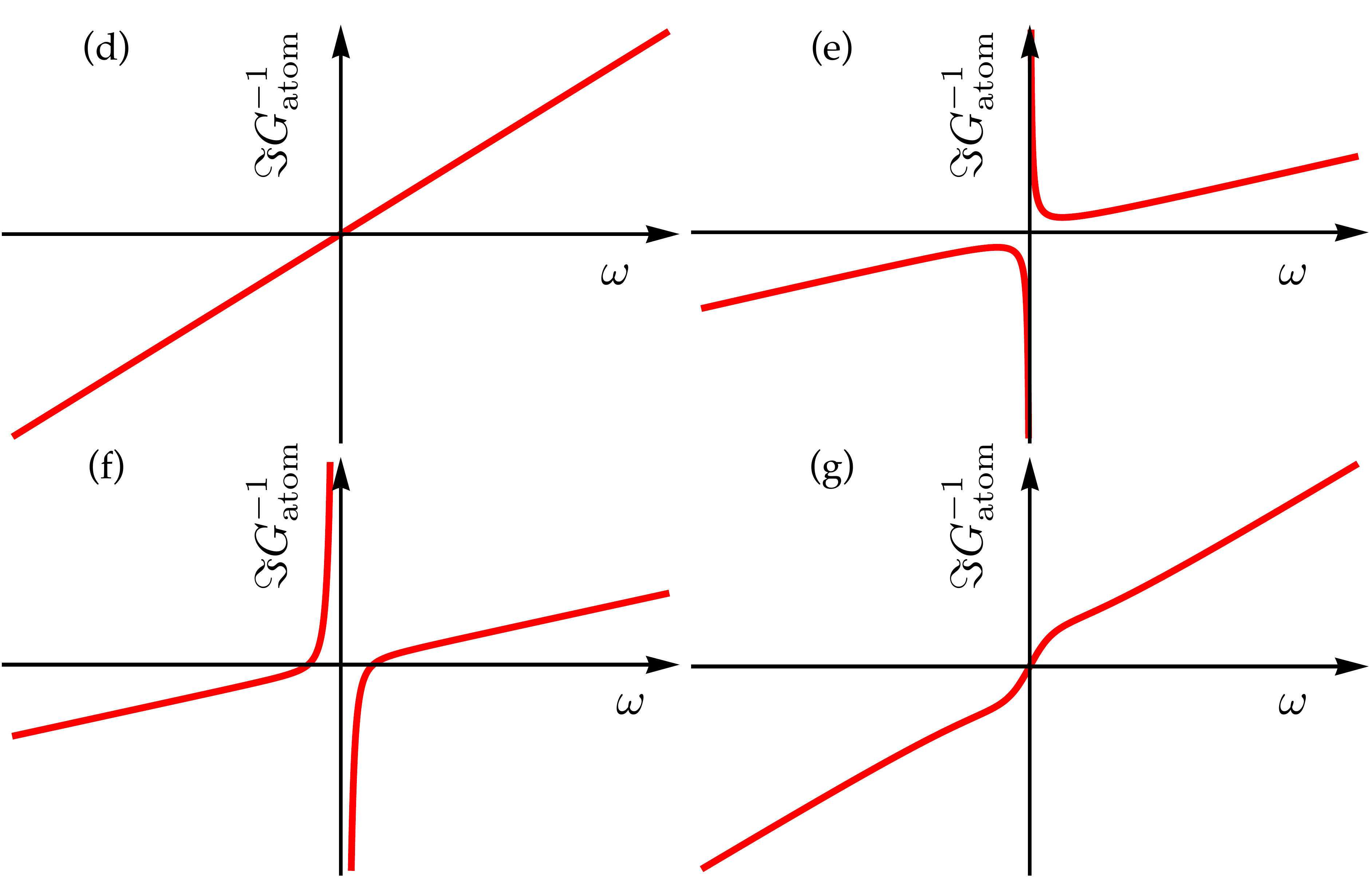}
   \caption{(a-c): The mapping $\omega\mapsto G^{-1}_\mathrm{atom}(\omega)$ appended with appropriate large circles defines a contour in the complex plane. The result of the contour integration gives FDWN. Cross denotes the singular points of the integrand $\frac{1}{(z^2-h_{\bold k}^{2})^{3}}$. Double line means the contour wraps around twice.  (d-g): The imaginary part of the PHS (d) $G^{-1}_\mathrm{atom}=$ $i\omega$,  (e) $i\omega-\frac{1}{i\omega}$,  (f) $i\omega-\frac{1}{(i\omega)^{3}} $ , (g) $i\omega - \frac{1}{i\omega-1} -\frac{1}{i\omega+1} $. They have FDWNs $1, 0, 2, 1$ respectively. The result could be read off from the number of right and left moving intersections of the curve with a horizontal line: $\gamma$ equals to the difference of them. One could also get the results from the complex plane contour integrations with contours drawn in (a),(b),(c),(a) respectively.}
   \label{fig:wrap}
\end{figure}

In the second step we appended appropriate large circles to render the integration on a closed contour $\mathcal{C}$, see Fig.\ref{fig:wrap}. Since $G_{\mathrm{atom}}^{- 1}$ is parameterized by $\omega$, the contour could be thought of as the path of $G_{\mathrm{atom}}^{- 1}(\omega)$ on the complex plane as $\omega$ varies from $-\infty$ to $\infty$. The result of the contour integration is determined by the singularities that contour $\mathcal{C}$ encloses. Notice that the residues at the two singularities $\mp|h_{\bold{k}}|$ are $\pm\frac{3}{16|h_{\bold{k}}|^{5}}$ respectively, we introduce $\gamma$ to denote their winding number differences \footnote[2]{One could proof that for a gapped system $\gamma$ is the same for momenta ${\bold k}$, otherwise, the contour $\mathcal{C}$ must cross the eigenvalues continuum of $H_{\bold k}$ and the system is thus gapless.}. Plug the result Eqn.\ref{eqn:freq} back into Eqn.\ref{eqn:separate}, we have

\begin{eqnarray}
n  =  \gamma \times \frac{3}{8 \pi^2} \int_{\mathrm{BZ}} \mathrm{d}^4k \,\varepsilon_{abcde} \hat{h}_{\bold{k}}^{a} \partial_{k_{x}} \hat{h}_{\bold{k}}^{b} \partial_{k_{y}} \hat{h}_{\bold{k}}^{c} \partial_{k_{z}} \hat{h}_{\bold{k}}^{d} \partial_{k_{\lambda}} \hat{h}_{\bold{k}}^{e}
\label{eqn:central}
\end{eqnarray}

Eqn. \ref{eqn:central} is the central result of the paper. FDWN $\gamma$ describes the winding of the atomic Green's function on the complex plane. It appears as a pre-factor of the momentum space TKNN number \cite{Qi:2008p12545}. In other words, the interacting Chern number $n$ is determined by the winding numbers in both the frequency and momentum spaces. Noninteracting case has $\gamma =1 $ since the frequency dependency of the noninteracting Green's function comes trivially from $i\omega$, see Fig.\ref{fig:wrap}(a). The presence of interacting induced self-energy would arguably give more complex contours and thus change the value of FDWN, see Fig.\ref{fig:wrap}(b-c). By this mechanism, local fluctuation effect changes the topological class through a multiplicity integer factor.

In case of the particle-hole symmetric (PHS) case, the meaning of FDWN is even simpler and appealing. In fact, with PHS the self-energy is purely imaginary and so does the atomic Green's function, \textit{i.e.} $ G_{\mathrm{atom}}^{-1}= i f(\omega)$ with $f$ a real function.  We redo the frequency integration in Eqn.\ref{eqn:separate},

\begin{widetext}

\begin{eqnarray}
\int_{- \infty}^{\infty} \frac{\mathrm {d} \omega }{2 \pi i}  \frac{\partial_{\omega}  G_{\mathrm{atom}}^{- 1}}{(G_{\mathrm{atom}}^{- 2} - |h_{\bold{k}}|^2)^3}
  & = &-\int_{f (- \infty)}^{f ( \infty)} \frac{\mathrm{d}f}{2 \pi}  {\frac{1}{(f^2 + |h_{\bold{k}}|^2)^3}} \nn
 & = & \frac{-3}{16\pi |h_{\bold{k}}|^5} [ \frac{2f/|h_{\bold{k}}|}{3 (1+ f^{2}/|h_{\bold{k}}|^{2})^2}+\frac{f/|h_{\bold{k}}|}{1+f^{2}/|h_{\bold{k}}|^{2}} +  {\arctan} ( \frac{f}{|h_{\bold{k}}|})] |^{f (\infty)}_{f (- \infty)}   \nn
& = & \frac{-3}{16 |h_{\bold{k}}|^5} \deg \{f\}
\end{eqnarray}

\end{widetext}

The contribution purely comes from the $\arctan$ functions.  Notice that $\omega \mapsto f(\omega)$ defines a mapping from $\mathbb{R}$ to $\mathbb{R}$, we could view $\pm\infty$ as the identical points and compactification the space $\mathbb{R}$ into a circle $S^1$ with stereographic projection. Then the function $f$ maps $S^{1}$ to $S^{1}$ and one has well defined winding numbers $\deg \{f\}$ of the mapping. One nice thing at the PHS case is the FDWN could be calculated by simply countering the right moving and left moving intersection of $f(\omega)$ with a given a horizontal line, see Fig.\ref{fig:wrap}.  For example, in the noninteracting case, $f(\omega)=\omega$ indicates $\gamma=1$, Fig.\ref{fig:wrap}(a). Nontrivial self-energies like $\Sigma(\omega)=\frac{1}{i\omega}$ ($f =\omega +\frac{1}{\omega}$) gives $\gamma=0$,  $\Sigma(\omega)=\frac{1}{(i\omega)^{3}}$  ($f =\omega -\frac{1}{\omega^{3}}$) results $\gamma=2$ and  $\Sigma(\omega)=\frac{1}{i\omega-1}+\frac{1}{i\omega+1}$ ($f=\omega+\frac{2\omega}{\omega^{2}+1}$) gives $\gamma=1$. The resulting FDWN are in accodance with the complex integration of the $G_{\mathrm{atom}}^{-1}$ in the complex plane (Fig.\ref{fig:wrap}(a-c)). These results are also been validated by direct numerical integrating Eqn.\ref{eqn:ishikawa}, see fellowing for discussions on the numerical details.

In general, the interacting Green's functions calculated from numerical solvers may have no analytical expression and have much complicated behaviors. The above rules still apply: one can simply read out the FDWN by plotting the Green's function and counting the number of intersections or winding numbers.

We then discuss some physical consequences of the FDWN. First, it indicates that the \emph{local fluctuations} can affect the topological properties of the system by developing nontrivial frequency dependency of the Green's functions. Second, through this mechanism, the topological phase could be broken \emph{without developing LRO}. This phenomenon has been uncovered in previous studies on the interacting Haldane model\cite{Wang:2010p25724}. Moreover, $\Sigma(\omega)=\frac{1}{i\omega}$ is actually corresponds to the Mott insulating states\cite{Georges:1993p715}. The fact of it has zero FDWN indicates that Mott insulating state in this sense has zero Chern number. To clarify potential misconceptions, the topological Mott insulator state refereed in \cite{Raghu:2008p3865} is actually a mean-field bond-order wave state with trivial FDWNs $\gamma = 1$. We would also like to remark that in many of the proposed interaction induced topological insulating phases \cite{Dzero:2010p15316, Wen:2010p20806, Sun:2011p35265}, interaction effects are mostly considered to give nontrivial momentum dependence of the mean-field self-energy $\Sigma_{\bold k}$, which is effectively absorbed into the noninteracting Hamiltonian $H_{\bold k}$.

\section{Three dimensional $Z_{2}$ topological insulators \label{sec:Z2}}
In this section we show that the FDWN is also relevant to the  $Z_{2}$ index of the three dimensional topological insulators. Ref.\cite{Qi:2008p12545, Wang:2010p25171} show that the three dimensional $Z_{2}$ topological insulators are closed related to the four dimensional Chern insulators through the dimension reduction. Given noninteracting Hamiltonian $H_{\bold k}$ (${\bold k}$ now denotes three dimensional momentum), the three dimensional $Z_{2}$ index could be calculated through a similar formula as Eqn.\ref{eqn:ishikawa}. We extend $H_{\bold k}$ to four dimension \cite{Li:2010p15108} by introducing a pumping parameter $k_{\lambda}$ and  $H_{\bold k} \rightarrow H_{\bold k} + k_{\lambda} \Gamma^{4}$ (The choice preserves the time reversal symmetry when $k_{\lambda}=\infty$, see \cite{Li:2010p15108}). Under the local self-energy approximation, we integrate over the frequency $\omega$ and the pumping parameter $k_{\lambda}$ and get,

\begin{widetext}

\begin{eqnarray}
n & =&  \frac{\varepsilon_{\mu \nu \rho\sigma\tau }}{15 \pi^2} \mathrm{Tr} \int_{\mathrm{BZ}}  \frac{\mathrm{d}^{3}k}{(2\pi)^{3}}  \int_{0}^{\infty}  \frac{dk_{\lambda}}{2\pi}   \int_{-\infty}^{\infty} \frac{\mathrm{d}\omega}{2\pi}  
 G \partial_{\mu} G^{- 1} G \partial_{\nu} G^{- 1} G \partial_{\rho} G^{- 1}G \partial_{\sigma} G^{- 1}G \partial_{\tau} G^{- 1}  \nn
  & = & \gamma \times \frac{3}{8 \pi^{2}} \int_{\mathrm{BZ}} \mathrm{d}^3 k  \int_0^{\infty} dk_\lambda  \frac{1}{|h_{\bold{k}}|^5} \varepsilon_{abcde} h_{\bold{k}}^{a} \partial_{k_{x}} h_{\bold{k}}^{b} \partial_{k_{y}}
  h_{\bold{k}}^{c} \partial_{k_{z}} h_{\bold{k}}^{d} \partial_{k_\lambda} h_{\bold{k}}^{e} \nn
  & = &\gamma \times \frac{1}{8 \pi^{2}} \int_{\mathrm{BZ}} \mathrm{d}^3 k \frac{2|h_{\bold{k}}| + h_{\bold{k}}^{4}}{(|h_{\bold{k}}|
  + h_{\bold{k}}^{4})^2 |h_{\bold{k}}|^3} \varepsilon_{abcd} h_{\bold{k}}^{a} \partial_{k_{x}} h_{\bold{k}}^{b} \partial_{k_{y}} h_{\bold{k}}^{c}
  \partial_{k_{z}} h_{\bold{k}}^{d}  \nn
   & = & \gamma \times \ell 
   \label{eqn:Z2}
\end{eqnarray}

\end{widetext}

The above formula shows that the FDWN $\gamma$ appears as a multiply factor with the momentum space $Z_{2}$ index $\ell$ (which is defined upon module $2$). As an example, the noninteracting $Z_{2}$ topological insulators has odd momentum space $Z_{2}$ index multiplied with $\gamma = 1$.

Notice that to get nontrivial (odd) $Z_{2}$ index, we need both $\gamma$ and $\ell$ to be an odd number. This leads to the conclusion that the only way FDWN plays a role of changing topological classes is to break down a noninteracting $Z_{2}$ topological insulators, \textit{i.e.}, there is no way to develop a topological nontrivial states from an trivial noninteracting state. (since even integer times any integer still gives an even integer numbers)

It is interesting to look at the three representative cases of the topological trivial states: (a). odd $\gamma$ even $\ell$, (b). even $\gamma$  even $\ell$ and (c). even $\gamma$ odd $\ell$. Physically, they correspond to (a). noninteracting band insulators, (b). Mott insulators and (c). a state with nontrivial momentum space topology, but overtakes by strong interactions induced even FDWN. Notice that in the last case it is not necessary to have LRO. It may be a candidate description of the spin liquid state uncovered in the studies of the KMH model\cite{Hohenadler:2011p28789, Yu:2011p35101, Zheng:2010p24293}. It is interesting to study the FDWN inside the spin liquid phase. See the following for discussions on the effect of the spatial fluctuations and practical calculations.



\section{Discussions \label{sec:practical}}

In this section, we comment on the practical calculation of the interacting Green's functions and the techniques for the frequency-momentum integration in Eqn.\ref{eqn:ishikawa}.

Firstly, under the local assumption, it's suitable to study the interaction effect on topological insulators using DMFT. Since the FDWN only depends on the overall shape of the self-energy, simple impurity solvers like IPT \cite{Georges:1993p715, Georges:1996p5571} will do the trick. And since only the Matsubara Green's function is needed, the advanced CTQMC solvers \cite{Rubtsov:2005p2260,Werner:2006p2175, Gull:2011p32237} could play major role in exact solution of the auxiliary impurity problems.

Secondly, under the local self-energy approximation, one could get similar results as Eqn.\ref{eqn:central} and and Eqn.\ref{eqn:Z2} for the two dimensional Chern and $Z_{2}$ insulators. However, one caveat is that the local self-energy assumption may be not valid in low dimensions. Nevertheless, the FDWN could still be the driving force for some of the interaction induced topological transitions. In case of the strong momentum dependence self-energies, diagrammatic calculations such as fluctuation-exchange approximation (FLEX) \cite{Bickers:1989p853} could be used to calculate the self-energy in the perturbative region. Another direction is using the dynamical cluster approximation \cite{Maier:2005p5551} or diagrammatic Monte Carlo methods \cite{Houcke:2010p29479} to incorporate the momentum dependence of the self-energies non-perturbatively. These results will tell how much the spatial fluctuations affect the results reported here. It's also interesting to see the synergies of these conventional many-body numerical methods with the study of topological insulators though the calculation of its interacting Green's functions.

Thirdly, we remark on the numerics for the integration of Eqn.\ref{eqn:ishikawa}. In cases such as the frequency and momentum dependence of the self-energy interwind or $\Sigma(\omega,{\bold k})$ comes from the above mentioned numerical methods, one need to perform the five-dimensional integration in Eqn.\ref{eqn:ishikawa} numerically. Since usually the contributions of the integrand are concentrated around several particular points in the frequency-momentum space ($\omega=0$ and several special k-points), it advised to use Monte Carlo technique to sample the integrand. In our case, we find the VEGAS algorithm \cite{Lepage:1978p26048} (which has been used for more than thirty years in the high-energy physics for calculating multi-dimensional quadrature) is particularly suitable for this kind of integration.

Finally, for realistic topological insulating materials which are interacting, it maybe possible to read out the frequency-momentum dependency of the self-energy from the ARPES data and plug into Eqn.\ref{eqn:ishikawa} to tell whether the the material is in the topological phase or not.

\section{Conclusions}

In conclusion, we studied the interaction effect on topological insulators through their single-particle Green's functions. Under the local self-energy approximation, we uncover a mechanism where the winding in the frequency domain changes the topological classes. FDWN provide a mechanism of breaking down the topological phases without developing LRO. The finding is in accordance with the ongoing numerical studies on the interacting topological insulators \cite{Hohenadler:2011p28789, Yu:2011p35101, Zheng:2010p24293, Wang:2010p25724}. We urge the interpretation of the featureless intermediate phases uncovered in these studies based on the FDWN.


\section{Acknowledgment}
The work is supported by NSF-China and MOST-China. XCX is also supported in part by DOE through DE-FG02-04ER46124. LW thanks Hua Jiang, Zhong Wang, Yuan Wan and Philipp Werner for helpful discussions.

\bibliography{/Users/wanglei/Documents/Papers/papers}
\end{document}